\documentstyle[amssymb,12pt,epsfig]{article}
\newlength{\dinwidth}
\newlength{\dinmargin}
\begin{document}

%--------------------------------------------------------------------------

%--------------------------------------------------------------------------
%--------------------------------------------------------------------------

\thispagestyle{empty}  \rightline{Napoli DSF-T-39/2001} %
\rightline{INFN-Na-39/2001}
\rightline{cond-mat/0204575} \vspace{1cm}

\begin{center}
{\LARGE {\bf Paired states on a torus} }

\vspace{8mm}

{\large Gerardo Cristofano, Giuseppe Maiella, \ }

{\large Vincenzo Marotta, Giuliano Niccoli } \vspace{1cm}

{\it Dipartimento di Scienze Fisiche \\[0pt]
Universit\'{a} di Napoli ``Federico II'' \\[0pt]
and \\[0pt]
INFN, Sezione di Napoli }

{\small Via Cinthia - Compl.\ universitario M. Sant'Angelo - 80126 Napoli,
Italy }

{\bf Abstract\\[0pt]
}
\end{center}

\begin{quotation}
We analyze the modular properties of the effective CFT description for
paired states, proposed in \cite{cgm2}, corresponding to the non-standard
filling $\ \nu =\frac{1}{p+1}$. We construct its characters  for the twisted
and the untwisted sector and the diagonal partition function. We show that
the degrees of freedom entering our partition function naturally go to
complete a $Z_{2}$-orbifold construction of the CFT for the Halperin state.
Different behaviours for the $p$ even and $p$ odd cases are also studied.
Finally it is shown that the tunneling phenomenon selects out a twist
invariant CFT which is identified with the Moore-Read model.

\vspace*{0.5cm}



{\footnotesize Work supported in part by the European Communities Human
Potential}

{\footnotesize Program under contract HPRN-CT-2000-00131 Quantum Spacetime}

\vfill
{\small {\bf
\begin{tabbing}
-------------------------------------------------------------\\
\=  gerardo.cristofano@na.infn.it \\
\=  giuseppe.maiella@na.infn.it    \\
\=  vincenzo.marotta@na.infn.it   \\
\=  giuliano.niccoli@na.infn.it
\end{tabbing}}}\newpage \baselineskip=18pt \setcounter{page}{2}
\end{quotation}

\section{Introduction}

Recently an Effective Conformal Field Theory (ECFT) of the so-called paired
and pa\-ra\-fer\-mio\-nic Quantum Hall (QH) states \cite{R-R}, at fillings $%
\nu =\frac{m}{pm+2}$, describing a new class of universality of non standard
QH plateaux has been proposed \cite{cgm2}. Such a CFT\ is built out of $m$
scalar fields, one describing the charged modes and the remaining $m-1$
ones, which satisfy twisted boundary conditions, describe the so called
neutral modes. The interesting (and new) aspect of the construction employed
(the so called $m$-reduction procedure) for both the Jain fillings $\nu =%
\frac{m}{pm+1}$ and the non standard ones, were analyzed in \cite{cgm1} and
\cite{cgm2} where it was shown how naturally it induces the twist on the
boundary, given by a $Z_{m}$ group generated by the phases $\epsilon ^{j}=e^{%
\frac{2\pi ij}{m}}$, $j=1,...,m$. As a consequence the chiral primary fields
appear as composite operators with a charged and a neutral component. It has
been proved that their correlators reproduce the ground state wave function
for the Jain as well as for the paired states, generalizing the Laughlin
type wave function on the plane \cite{cgm2},\cite{cgm1}. The role of the
neutral components described by the untwisted vertices is to cancel out the $%
m$-root singularity present in the correlators of the charged ones. On a
physical ground the neutral component describes the interaction between the
electrons of the $m$ Landau levels (or layers) involved in the formation of
the ground and excited states. In this paper we present the detailed
analysis of such a twisted CFT for the topology of the torus (genus $g=1$
Riemann surface).

First we construct the characters associated to the chiral vertices (already
known for the theory on the plane \cite{cgm2}), their modular properties
(under $SL(2,Z)$ transformations or one of its subgroups), and finally we
exhibit the partition function. All that allows us to develop a
classification scheme of these twisted (orbifold) CFTs. A very interesting
case seems to be the possibility to select out a subclass of characters by
imposing the invariance under the twist group, which defines a CFT with a
smaller central charge but with a higher symmetry.

These theories are related to the models introduced some years ago in \cite
{M-R},\cite{R-R} which has been also obtained in \cite{Cappelli-geo-todo}
\cite{frohlich} as a coset of $c=2m$ CFTs. The relevant aspect in our
construction of such a rational CFT, which from now on we will refer to
Twisted Model (TM),\ is the presence of composite operators in which the $%
Z_{m}$ (center of $SU(m)$ symmetry of the untwisted vertices) couples to the
$Z_{m}$ subgroup of $U(1)$ relative to the charged vertices.

As it is well known, for the applications to the QH fluid physics the
general properties of the one point functions on the torus and of the
characters give the most significative information on the topological
aspects of the QH physics (and also on the edge deformations), as the
degeneracy of the ground state wave function on the torus, the braiding
properties or generalized statistics \cite{Stone}, the conductance $\sigma
_{H}$, which defines the so called ``topological order'' \cite{wen}\cite
{cgm5}. One of the new phenomena induced by the topological properties is an
effective strong coupling between the layers. For the particular $m=2$ case,
that is for a two layers system, corresponding to an effective filling $\nu =%
\frac{1}{p+1}$, such a phenomenon, found experimentally in \cite{Sp-}, was
analyzed in \cite{cgm3} within our CFT description and in the context of two
interacting 2D-branes. Basically that phenomenon has its origin in the
topology of the $m$-covering ($m=2$) of the torus, as we will see in this
paper. Then we have all the ingredients for confronting the properties of
the TM partition function under the infinite discrete symmetry $SL(2,Z),$ or
its subgroups $\Gamma _{0}(2)$, for the $c=2$ case, with the $331$ model and
its generalization ( the ($2+p$,$2+p$,$p$) Halperin series (H) \cite
{Halperin0}) and only in part with the Haldane-Rezayi model (HR) \cite
{Haldan-R}. The sub-theory with $c=\frac{3}{2}$ proposed by Moore and Read
(MR) in \cite{M-R} for reproducing the Pfaffian and its qPfaffian
generalization is also analyzed on the torus, confirming the presence of a
new deeper symmetry i.e. the $N=2$ superconformal one for the special $p=1$
case.

The paper is organized as follows:

In sec. 2 we discuss the ground state wave functions on the plane of a two
interacting layers system which have been first analyzed in \cite{Halperin0}%
, \cite{M-R} and \cite{Haldan-R}.

In sec. 3 we briefly summarize the $m$-reduction construction on the plane
(genus $g=0$), which has been used in \cite{cgm1} to construct CFT models
for the Jain series at filling $\nu =\frac{m}{pm+1}$ and in \cite{cgm2} for
the parafermionic serie at fillings $\nu =\frac{m}{pm+2}$. Then we give the
primary fields and their correlation functions for the $m=2$ case,
evidencing the simple relation of the H states with our proposed ground
state for the plane geometry \cite{cgm2}.

In sec. 4 we generalize the $m$-reduction\ construction of the TM for the
torus topology; then we derive the conformal blocks both for the twisted and
the untwisted sectors. Finally we briefly comment on the possible transition
from Abelian to non Abelian statistics \cite{M-R},\cite{Naiack}.

In sec. 5 we construct the complete diagonal partition function for this CFT
with central charge $c=2$ following the scheme of secs. 2, 3 and 4. We prove
that for the $p$ even case, for which boson like pseudoparticles (or
excitations) are formed, it is $SL(2,Z)$ invariant while for the $p$ odd
case, for which fermions-like excitations are formed, it is only invariant
under the $\Gamma _{0}(2)$ subgroup of $SL(2,Z)$. For p even a comparison
with the $Z_{2}$-orbifold construction is given.

In the conclusions (sec. 6) a comparison of the TM with the H and MR models
is given. Further comments on the problem of the non-Abelian statistics are
briefly addressed together with a plausible physical interpretation of the
twisted boundary conditions.

In App. A the characters for the generic $m$ case are explicitly constructed
using the $m$-reduction procedure on the torus; then their factorization in
terms of the charged and neutral components is derived.

In App. B the explicit derivation of the modular transformation properties
are given for the $m=2$ case.

\section{Bilayer Quantum Hall System}

We are interested in a system consisting of two parallel layers of 2D
electrons gas in a strong perpendicular magnetic field B.

The states which will be considered are the Halperin states \cite{Halperin0}%
, the Haldane-Rezayi \cite{Haldan-R} and the Moore-Read states \cite{M-R}.
The simplest abelian quantum Hall state in a disc topology is described by a
generalization of the Laughlin wave function first introduced by Halperin
\cite{Halperin0}:
\begin{equation}
f^{H}(\{z_{i}^{(a)}\})=\prod_{i<j}\left( z_{i}^{(1)}-z_{j}^{(1)}\right)
^{2+p}\prod_{i<j}\left( z_{i}^{(2)}-z_{j}^{(2)}\right)
^{2+p}\prod_{i,j}\left( z_{i}^{(1)}-z_{j}^{(2)}\right) ^{p}e^{-\frac{1}{4}%
\sum_{i,a}|z_{i}^{(a)}|^{2}}
\end{equation}
where $z_{i}^{(a)}$ is the complex coordinate of the electron $i$ in the
layer $(a)$.

Here $p$ is a positive integer, odd (for the fermionic series) or even (for
the bosonic one), characterizing the flux attached to the particles. These
states are incompressible and their gapless excitations are confined to the
droplet edges, which have length L and are parametrized by the coordinate $%
x_{i}^{(a)}$ ( $z_{i}^{(a)}=e^{i\frac{x_{i}^{(a)}}{2\pi L}}$). The filling
factor $\nu ^{(a)}=\frac{1}{2p+2}$ is the same for the two layers (balanced
system) while the total filling is $\nu =\nu ^{(1)}+\nu ^{(2)}=\frac{1}{p+1}$%
. Halperin series contains, in particular, the fermionic $331$ state which
corresponds to $p=1$ and the bosonic $220$ state corresponding to $p=0$. The
CFT description for these models can be given in terms of two compactified
chiral bosons $Q^{(a)}$ with central charge $c=2$. An equivalent description
is given in terms of fields which diagonalize the interlayer interaction $X$
, $\phi $ given by: $X=Q^{(1)}+Q^{(2)}$, $\phi =Q^{(1)}-Q^{(2)}$, where $%
Q^{(a)}$, $a=1,2$ are defined on a single layer.

Then for the bilayer we are considering $X$ carries the total charge with
velocity v$_{X},$ while $\phi $ carries the charge difference of the two
edges with velocity $v_{\phi }$ i.e. no charge, being the number of
electrons the same for each layer \cite{priadko}. We refer to $X$ and $\phi $
as to the charged and neutral field, respectively and restrict our
discussion to the states in which both edge modes move in the same direction.

In such a basis the wave function is given by
\begin{equation}
f^{H}(\{z_{i}^{(a)}\})=f_{C}(\{z_{i}^{(a)}\})f_{N}^{H}(\{z_{i}^{(a)}\})e^{-%
\frac{1}{4}\sum_{i,a}|z_{i}^{(a)}|^{2}}
\end{equation}
where
\begin{equation}
f_{C}(\{z_{i}^{(a)}\})=\prod_{i<j}\left( z_{i}^{(1)}-z_{j}^{(1)}\right)
^{p+1}\prod_{i<j}\left( z_{i}^{(2)}-z_{j}^{(2)}\right)
^{p+1}\prod_{i,j}\left( z_{i}^{(1)}-z_{j}^{(2)}\right) ^{p+1}
\end{equation}
and
\begin{equation}
f_{N}^{H}(\{z_{i}^{(a)}\})=Det\left( \frac{1}{z_{i}-z_{j}}\right)
=(-1)^{N(N-1)/2}\frac{\prod_{i<j}\left( z_{i}^{(1)}-z_{j}^{(1)}\right)
\prod_{i<j}\left( z_{i}^{(2)}-z_{j}^{(2)}\right) }{\prod_{i,j}\left(
z_{i}^{(1)}-z_{j}^{(2)}\right) }
\end{equation}
are the holomorphic wave function of the charged and neutral modes.

Let us notice that for the H model there is no interlayer current, therefore
the fields $Q^{(a)}$ are defined on a single layer only and there is no
connection between them, i.e. the two edges are not ``connected''. Instead
for the case of the MR model or, in general, for the twisted sector of the
TM that is not the case.

The other ground state wave functions proposed for the bilayer system differ
one from another only in their neutral degrees of freedom contribution. They
are given by:
\begin{equation}
f_{N}^{HR}(\{z_{i}^{(a)}\})=Det\left( \frac{1}{\left( z_{i}-z_{j}\right) ^{2}%
}\right)
\end{equation}
for the HR state, where $Det$ is the determinant and
\begin{equation}
f_{N}^{MR}(\{z_{i}^{(a)}\})=Pf\left( \frac{1}{\left( z_{i}-z_{j}\right) }%
\right)
\end{equation}
for the MR case, where $Pf\left( \frac{1}{z_{i}-z_{i^{\prime }}}\right) =%
{\cal A}\left( \frac{1}{z_{1}-z_{2}}\frac{1}{z_{3}-z_{4}}\dots \right) $ is
the antisymmetrized product over pairs of electrons or equivalently the
square root of the determinant.

The neutral sector of the edge excitations for MR, HR and H states are given
by Majorana-Weyl, symplectic and Dirac fermion respectively (while in the
model here analyzed we have (Ising)$^{2}$). They can be seen as the QH
analogues of a BCS superconductor with a would be order parameter
(associated to the pairing phenomenon) which is $S_{z}=1,0$ $p$-wave for MR
and H respectively and $d$-wave for the HR. From a CFT point of view MR has
central charge $c=3/2$ while the H model has $c=2$ and HR is the (unitary)
projection of a $c=-1$ (its peculiar properties will be not discussed here).
Beside the relevant physical differences between the three states just seen,
for all cases the charged and neutral sectors are coupled by a $Z_{2}$
(parity rule) symmetry \cite{Cappelli-geo-todo}, \cite{INO} (as a
consequence of the monodromy property for the CFT); we notice that in the
methods used in \cite{cgm2} such a selection rule is built in by
construction and was called the $m$-ality constraint.

The edge excitations are generated by acting with the extended chiral
algebra on the Highest Weight States (HWS). These fields also correspond to
quasi-particles in the bulk. The complete description of the edge
excitations are conveniently summarized in the extended chiral algebra
characters $\chi (w|\tau )$. By taking $\tau =-iT_{0}/k_{B}T$ ,\ where $k_{B}
$ is the Boltzman constant and $T_{0}=\hslash v_{F}/L$ is the level spacing
of the system, combinations of characters represent contributions from each
sector to the grand partition function.

On the other hand in the TM (or abelian $Z_{2}$ orbifold) the boundary
conditions connect the two scalar chiral fields $Q^{(a)}$ which can be
thought as components of a unique ``boson'' defined on a double covering of
the disc ($z_{i}^{(1)}=-z_{i}^{(2)}=z_{i}$) see Fig. 1.
\begin{figure}[h]
\centerline{\epsfxsize=2cm} \epsffile{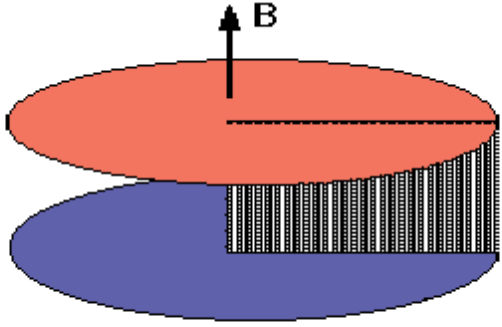}
\caption{Twisted bilayer system}
\label{fig:twist}
\end{figure}
That is in strong contrast with the Halperin model. Correspondingly also the
radius gets renormalized and the contributions for the charged and neutral
components are given by:
\begin{eqnarray}
f_{C}(\{z_{i}\}) &=&\prod_{i<j}\left( z_{i}^{2}-z_{j}^{2}\right)
^{p+1}\prod_{i}\left( 2z_{i}\right) ^{\frac{p+1}{2}}  \label{twistwave} \\
f_{N}^{TM}(\{z_{i}\}) &=&(-1)^{N(N-1)/2}\frac{\prod_{i<j}\left(
z_{i}-z_{j}\right) ^{2}}{\prod_{i<j}\left( z_{i}^{2}-z_{j}^{2}\right) }%
\prod_{i}\left( 2z_{i}\right) ^{\frac{1}{2}}
\end{eqnarray}
It is easy to check that, by using the map $z_{i}^{2}\rightarrow z_{i}$ we
obtain the wave function given in \cite{cgm2} (without the symmetrization).
The two layers system becomes equivalent to a one-layer QHF with filling $%
\nu =1/(p+1)$ with interacting spin degrees of freedom. Nevertheless, the
neutral sector which describes this interaction, becomes an (Ising)$^{2}$
model due to the decoupling of the spin structures \cite{cft}.

\section{The m-reduction on the plane}

In this section we briefly review the $m$-reduction procedure on the plane
as a starting point for its generalization to the torus which is the main
content of this paper. Our approach is meant to describe all the plateaux
with even denominator starting from the bosonic Laughlin filling $\nu
=1/(pm+2)$, which is described by a CFT with $c=1$, in terms of a scalar
chiral field compactified on a circle with radius $R^{2}=1/\nu =pm+2$ (or
its dual $R^{2}=4/(pm+2)$). Then the $U(1)$ current is given by $%
J(z)=i\partial _{z}Q(z)$, where $Q(z)$ is the compactified Fubini field with
the standard mode expansion:
\begin{equation}
Q(z)=q-i\,p\,lnz+\sum_{n\neq 0}\frac{a_{n}}{n}z^{-n}  \label{modes}
\end{equation}
with $a_{n}$, $q$ and $p$ satisfying the commutation relations $\left[
a_{n},a_{n^{\prime }}\right] =n\delta _{n,n^{\prime }}$ and $\left[ q,p%
\right] =i$.

The primary fields are expressed in terms of the vertex operators $U^{\alpha
}(z)=:e^{i\alpha Q(z)}:$ with $\alpha ^{2}=1,...,2+pm$ and conformal
dimension $h=\frac{\alpha ^{2}}{2}$.

We start with this set of fields in the above CFT (mother theory). Using the
$m$-reduction procedure, which consists in considering the subalgebra
generated only by the modes in eq.(\ref{modes}), which are a multiple of an
integer $m$, we get the image of the twisted sector of a $c=m$ orbifold CFT
(daughter theory, i.e. the TM) which describes the LLL dynamics.

Then the fields in the mother CFT can be factorized into irreducible orbits
of the discrete $Z_{m}$ group which is a symmetry of the TM and can be
organized into components which have well defined transformation properties
under this group. To compare the orbifold so built with the $c=m$ CFT, we
use the mapping $z\rightarrow z^{1/m}$and the isomorphism defined in ref.
\cite{VM} between fields on the $z$ plane and fields on the $z^{m}$ covering
plane given by the following identifications: $a_{nm+l}\longrightarrow \sqrt{%
m}a_{n+l/m}$, $q\longrightarrow \frac{1}{\sqrt{m}}q$.

It is useful to define the invariant scalar field
\begin{equation}
X(z)=\frac{1}{m}\sum_{j=1}^{m}Q(\varepsilon ^{j}z)  \label{X}
\end{equation}
where $\varepsilon ^{j}=e^{i\frac{2\pi j}{m}}$, corresponding to a
compactified boson on a circle with radius now equal to $%
R_{X}^{2}=R^{2}/m=p+2/m$. This field describes the $U(1)$ electrically
charged sector of the new filling.

On the other hand, the non-invariant fields defined by
\begin{equation}
\phi ^{j}(z)=Q(\varepsilon ^{j}z)-X(z),~~~~~~~~\ \ \ \ \ \ \ \ \ \
~~~~~~~~\sum_{j=1}^{m}\phi ^{j}(z)=0  \label{phi}
\end{equation}
naturally satisfy twisted boundary conditions and the $J(z)$ current of the
mother theory decomposes into a charged current given by $J(z)=i\partial
_{z}X(z)$ and $m-1$ neutral ones $\partial _{z}\phi ^{j}(z)$ \cite{cgm2}\cite
{PS}.

In the same way every vertex operator in the mother theory can be factorized
in a vertex that depends only on the invariant field:
\begin{equation}
{\cal U}^{\alpha }(z)=z^{\frac{\alpha ^{2}}{2}\frac{(m-1)}{m}}:e^{i\alpha {{%
\ {{\cdot} }}}X(z)}:~~\ \ \ \ \ \ \ \ ~\alpha ^{2}=1,...,2+pm
\end{equation}
and in vertex operators depending on the $\phi ^{j}(z)$ fields.

We also introduce the neutral component:
\begin{equation}
\psi _{1}(z)=\frac{z^{\frac{(1-m)}{m}}}{m}\sum_{j=1}^{m}\varepsilon
^{j}:e^{i\alpha {{\cdot} }\phi ^{j}(z)}:  \label{par.1}
\end{equation}
which is twist invariant and generates the parafermionic OPE (see \cite{cgm2}
for details) and
\begin{equation}
\bar{\psi}_{g}(z)=\frac{z^{\frac{(1-m)}{m}}}{m}\sum_{j=1}^{m}\varepsilon
^{gj}:e^{i\alpha {{\cdot} }\phi ^{j}(z)}:
\end{equation}
with $g=2,...,m$, which are Z$_{m}$ twisted.

From now on we fix $m=2$ which is appropriate for the bilayer system.

From eq.(\ref{X}) and (\ref{phi}) the untwisted field $X$ is given by $X(z)=%
\frac{1}{2}\left( Q(z)+Q(-z)\right) $ and the twisted one by $\phi (z)=\frac{%
1}{2}\left( Q(z)-Q(-z)\right) $ which satisfies the boundary conditions $%
\phi (e^{\pi i}z)=-\phi (z)$ and describes the neutral sector.
Correspondingly, the Virasoro generator splits in two terms \cite{VM} both
contributing with $c=1$ to the central charge. They are:
\begin{equation}
T_{X}(z)=-{\frac{1}{2}}\left( \partial _{z}X(z)\right) ^{2}  \label{STRESSX}
\end{equation}
and
\begin{equation}
T_{\phi }(z)=-\frac{1}{4}\partial _{z}\phi (z)^{2}+\frac{1}{16z^{2}}
\label{STRESSFI}
\end{equation}
The primary fields are the composite operators $V(z)={\cal U}_{X}(z)\psi (z)$
where
\begin{equation}
{\cal U}_{X}(z)=\frac{1}{\sqrt{z}}:e^{i\alpha X(z)}:
\end{equation}
are the vertices of the charged sector with $\alpha ^{2}=2(p+1)$. The
neutral component
\begin{equation}
\psi (z)=\frac{1}{\sqrt{z}}:e^{i\alpha \phi (z)}:
\end{equation}
is built out of the twisted scalar field \cite{cgm2}. More explicitly, the
HWS of the neutral sector can be classified by two kinds of chiral operators:

the one which does not change the boundary conditions
\begin{equation}
\psi (z)=\frac{1}{2\sqrt{z}}\left( e^{i\alpha {{\cdot} }\phi (z)}+e^{i\alpha
{{\cdot} }\phi (-z)}\right)  \label{even}
\end{equation}

and the other which does
\begin{equation}
\bar{\psi}(z)=\frac{1}{2\sqrt{z}}\left( e^{i\alpha {{\cdot} }\phi
(z)}-e^{i\alpha {{\cdot} }\phi (-z)}\right) .  \label{odd}
\end{equation}
In the fermionized version one can see that they correspond to $c=1/2$
Majorana fermions with periodic (Ramond) or anti-periodic (Neveu-Schwarz)
boundary conditions \cite{cgm4}. We remind that the Ising model contains
three independent characters which can be chosen as \{$I,\psi ,\sigma $\} or
\{$I,\psi ,\mu $\}, where $\sigma $ and \ $\mu $ are both twist fields of
dimension $h=1/16$. They are also called order ($\sigma $) and disorder ($%
\mu $) fields and are non local with respect to each other (see for example
\cite{Ginsparg}).

The TM decomposes into a tensor product of two CFTs, a twisted invariant one
with $c=3/2$ while the second one has $c=1/2$ and is realized in terms of a
Majorana fermion in the twisted sector. Such a factorization will be even
more evident in the construction of the partition function.

When the spacing between the layers is kept fixed at a the value such that
the tunneling amplitude is large enough so as the two species of electrons
become indistinguishable the H states is predicted to flow to the MR states.
In \cite{Cabra} it was shown that in the presence of tunneling the two
Majorana fermions behave as free fields, but acquire different velocities
which are determined by the bare velocity of the neutral boson $v_{\phi }$\
and by the tunneling amplitude. When the velocity of one Majorana fermion
becomes zero the theory reduces to the $c=3/2$ CFT. The operator responsible
of this tunneling effect was identified as a cosine operator for $R^{2}=4$
\cite{priadko} (the total energy in the (Ising)$^{2}$ model \cite
{difrancesco}).

What is striking is that for the $Z_{2}$ even under twist theory a cosine
term automatically appears in the energy-momentum tensor of the part of the
neutral sector described by the Ramond fields. They are the degrees of
freedom which survive after the tunnelling effect has taken place and the $%
Z_{2}$ symmetry which exchanges the two Ising fermions is broken. Indeed one
gets for the bosonized energy-momentum tensor
\begin{equation}
T_{\psi }(z)=-\frac{1}{4}(\partial \phi )^{2}-\frac{1}{16z^{2}}\cos (2\sqrt{2%
}\phi )
\end{equation}
Thus the new stable vacuum is the ground state of a $c=3/2$ CFT. Such a
result, in our opinion, is a strong confirmation of our CFT approach to the
bilayer system.

We are now ready to give the holomorphic part of the ground state wave
function for the filling $\nu =1/(p+1)$. To such an extent we consider the $%
2N_{e}$ electrons correlator which factorizes into a Laughlin-Jastrow type
term coming from the charged sector:
\begin{equation}
<2N_{e}\alpha |\prod_{i=1}^{2N_{e}}{\cal U}_{X}(z_{i})|0>=\prod_{i<i^{\prime
}=1}^{2N_{e}}(z_{i}-z_{i^{\prime }})^{p+1}  \label{eq: UN}
\end{equation}
and a contribution coming from the neutral excitations:
\begin{equation}
<0|\prod_{i=1}^{2N_{e}}\psi (z_{i})|0>=\frac{(\sqrt{z_{i}}-\sqrt{z_{j}})^{2}%
}{\prod_{i<j}(z_{i}-z_{j})}  \label{eq: FUN}
\end{equation}
(Notice that here the vacuum state for the neutral modes is a twisted one).

For $p$ an odd integer (i.e. for $\nu =1/q$, $q=p+1$) we get for the
correlator of $2N_{e}$ electrons:
\begin{equation}
<2N_{e}\alpha |\prod_{i=1}^{2N_{e}}V^{\sqrt{2q}}(z_{i})|0>=\prod_{i<i^{%
\prime }=1}^{2N_{e}}(z_{i}-z_{i^{\prime }})^{q}Pf\left( \frac{1}{%
z_{i}-z_{i^{\prime }}}\right)
\end{equation}
which reproduces the wave function $f^{MR}$ \cite{M-R}.

In a similar way we also are able to evaluate correlators of $2N_{e}$
electrons in the presence of quasi-hole excitations \cite{cgm2}.

\section{The theory on the torus}

It is possible to apply the $m$-reduction technique to the topology of the
torus and obtain the conformal blocks of the TM out of the $U(1)$ conformal
blocks of the mother theory. This is extensively worked out for general $m$
in App. A. For the special $m=2$ case, we are interested in, the characters
of the TM are explicitly given by (see eq.(\ref{Chrel})):
\begin{equation}
\chi _{((s,i),\tilde{f})}(w|\tau )=\sum_{l=0}^{1}N_{{\large (}l,(q,\tilde{f})%
{\large )}}(\tau )K_{2(p+1)l+q}(w|\tau )
\end{equation}
where $q=2s+i$, $s=0,...,p$ and $(i,\tilde{f})=0,1$. The $c=1$ conformal
blocks given above are explicitly written in terms of the standard Jacobi $%
\Theta $ and the Dedekind $\eta $ functions as:
\begin{equation}
K_{2(p+1)l+q}(w|\tau )=\frac{1}{\eta (\tau )}\Theta \left[
\begin{array}{c}
\frac{2(p+1)l+q}{4\left( p+1\right) } \\
0
\end{array}
\right] \left( 2\left( p+1\right) w|4\left( p+1\right) \tau \right)
\label{charge}
\end{equation}
They describe the degrees of freedom of the charged component with radius $%
R_{X}^{2}=\frac{2p+2}{2}${\LARGE \ }and are in number of $4(p+1)$\ as it is
well known.

On the other hand the characters of the neutral sector are given by
\begin{equation}
N_{{\large (}l,(q,\tilde{f}){\large )}}(\tau )=\frac{1}{2}\sum_{j=0}^{1}e^{-%
\frac{2\pi i}{2}j(\tilde{f}+\frac{q^{2}}{4}-\frac{1}{24}-\frac{\left[
2(p+1)l+q\right] ^{2}}{4})}\frac{\eta (\tau )}{\eta (\frac{\tau +j}{m})}
\end{equation}
They can be factorized in terms of an (Ising)$^{2}$ i.e. $\chi _{0}$ (the
identity), $\chi _{\frac{1}{2}}$ (the energy) and $\chi _{\frac{1}{16}}$
(the twist) operators. Their explicit expressions are given in App. A. We
distinguish the two Ising sectors as follows:\ the twist invariant Ising
characters are unbarred while the non invariant ones are barred.

The characters which are even under the $Z_{2}$ twist of the boson $\phi $
determine a $c=3/2$ \ CFT and are explicitly given by:
\begin{eqnarray}
\chi _{(0,s)}^{MR}(w|\tau ) &=&\chi _{0}(\tau )K_{2s}\left( w|\tau \right)
+\chi _{\frac{1}{2}}(\tau )K_{2(p+s)+2}\left( w|\tau \right)  \\
\chi _{(1,s)}^{MR}(w|\tau ) &=&\chi _{\frac{1}{16}}(\tau )\left(
K_{2s+1}\left( w|\tau \right) +K_{2(p+s)+3}\left( w|\tau \right) \right)  \\
\chi _{(2,s)}^{MR}(w|\tau ) &=&\chi _{\frac{1}{2}}(\tau )K_{2s}\left( w|\tau
\right) +\chi _{0}(\tau )K_{2(p+s)+2}\left( w|\tau \right)
\end{eqnarray}
They are equal to the characters of the MR states \cite{M-R} and add up to $%
3(p+1)$. The characters of the $c=2$ CFT which describe the H states can be
also written in terms of these building blocks as follows:
\begin{eqnarray}
\chi _{(1,s)}^{H}(w|\tau ) &=&\bar{\chi}_{0}\chi _{(0,s)}^{MR}(w|\tau )+\bar{%
\chi}_{\frac{1}{2}}\chi _{(2,s)}^{MR}(w|\tau ) \\
\chi _{(2,s)}^{H}(w|\tau ) &=&\bar{\chi}_{0}\chi _{(2,s)}^{MR}(w|\tau )+\bar{%
\chi}_{\frac{1}{2}}\chi _{(0,s)}^{MR}(w|\tau ) \\
\chi _{(3,s)}^{H}(w|\tau ) &=&\chi _{(4,s)}^{H}(w|\tau )=\bar{\chi}_{\frac{1%
}{16}}\chi _{(1,s)}^{MR}(w|\tau )  \label{untwisted}
\end{eqnarray}
We see from the previous relations that the distinct characters for the H
model $\chi _{(3,s)}^{H}$ and \ $\chi _{(4,s)}^{H}$ are equal, on the other
hand, as we will see in the following, they correspond to an unique
character in the orbifold model. Notice also that for the H states the two
real fermions must have the same boundary conditions on the torus to
describe a Dirac field. Such a phenomenon could not be observed on the
plane, where the two fermions appeared totally decoupled.

\subsection{\label{twist}The characters of the twisted sector}

The characters of the twisted sector are denoted as $\chi _{(\left(
i,s\right) ,\tilde{f})}$, with the indices $i,$ $s,$ $\tilde{f}$ defined as
before and their explicit expression, which follows from eqs.$\left( \ref{le
b}\right) $, $\left( \ref{F(m,p)1}\right) $ and $\left( \ref{eq.11}\right) $
of App. A, depends on the parity of $p$:

for{\bf \ }$p${\bf \ }even
\begin{eqnarray}
\chi _{(\left( 0,s\right) ,0)}(w|\tau ) &=&\bar{\chi}_{\frac{1}{16}}\chi
_{(0,s)}^{MR}(w|\tau ) \\
\chi _{(\left( 0,s\right) ,1)}(w|\tau ) &=&\bar{\chi}_{\frac{1}{16}}\chi
_{(2,s)}^{MR}(w|\tau ) \\
\chi _{(\left( 1,s\right) ,0)}(w|\tau ) &=&\bar{\chi}_{0}\chi
_{(1,s)}^{MR}(w|\tau ) \\
\chi _{(\left( 1,s\right) ,1)}(w|\tau ) &=&\bar{\chi}_{\frac{1}{2}}\chi
_{(1,s)}^{MR}(w|\tau )
\end{eqnarray}
and add up to $4(p+1),$

for{\bf \ }$p${\bf \ }odd
\begin{eqnarray}
\chi _{((0,s),0)}(w|\tau ) &=&\bar{\chi}_{\frac{1}{16}}\chi _{0}\left(
K_{2s}\left( w|\tau \right) +K_{2(p+s)+2}\left( w|\tau \right) \right)  \\
\chi _{((0,s),1)}(w|\tau ) &=&\bar{\chi}_{\frac{1}{16}}\chi _{\frac{1}{2}%
}\left( K_{2s}\left( w|\tau \right) +K_{2(p+s)+2}\left( w|\tau \right)
\right)  \\
\chi _{((1,s),0)}(w|\tau ) &=&\chi _{\frac{1}{16}}\left( \bar{\chi}%
_{0}K_{2s+1}\left( w|\tau \right) +\bar{\chi}_{\frac{1}{2}%
}K_{2(p+s)+3}\left( w|\tau \right) \right)  \\
\chi _{((1,s),1)}(w|\tau ) &=&\chi _{\frac{1}{16}}\left( \bar{\chi}_{\frac{1%
}{2}}K_{2s+1}\left( w|\tau \right) +\bar{\chi}_{0}K_{2(p+s)+3}\left( w|\tau
\right) \right)
\end{eqnarray}
and add up to $4(p+1)$. If we define the symmetrized ($+$)\ and the
antisymmetrized ($-$) linear combinations, we see that the combinations $%
\chi _{(i,s)}^{+}$ do not depend on the parity of $p$ and are given by:
\begin{eqnarray}
\chi _{(0,s)}^{+}(w|\tau ) &=&\bar{\chi}_{\frac{1}{16}}\left( \chi
_{(0,s)}^{MR}(w|\tau )+\chi _{(2,s)}^{MR}(w|\tau )\right)  \\
\chi _{(1,s)}^{+}(w|\tau ) &=&\left( \bar{\chi}_{0}+\bar{\chi}_{\frac{1}{2}%
}\right) \chi _{(1,s)}^{MR}(w|\tau )
\end{eqnarray}
Instead the $\chi _{(i,s)}^{-}$ depend on the parity of $p$:

for $p$ even
\begin{eqnarray}
\chi _{(0,s)}^{-}(w|\tau ) &=&\bar{\chi}_{\frac{1}{16}}\left( \chi
_{(0,s)}^{MR}(w|\tau )-\chi _{(2,s)}^{MR}(w|\tau )\right) \\
\chi _{(1,s)}^{-}(w|\tau ) &=&\left( \bar{\chi}_{0}-\bar{\chi}_{\frac{1}{2}%
}\right) \chi _{(1,s)}^{MR}(w|\tau )
\end{eqnarray}

for $p$ odd
\begin{eqnarray}
\chi _{(0,s)}^{-}(w|\tau ) &=&\bar{\chi}_{\frac{1}{16}}\left( \chi _{0}-\chi
_{\frac{1}{2}}\right) \left( K_{2s}\left( w|\tau \right) +K_{2(p+s)+2}\left(
w|\tau \right) \right) \\
\chi _{(1,s)}^{-}(w|\tau ) &=&\chi _{\frac{1}{16}}\left( \bar{\chi}_{0}-\bar{%
\chi}_{\frac{1}{2}}\right) \left( K_{2s+1}\left( w|\tau \right)
-K_{2(p+s)+3}\left( w|\tau \right) \right)
\end{eqnarray}

For $p$ odd we need to take into account an extra symmetry due to the
fermionic nature of \ the particles while the starting theory is a bosonic
one. From a mathematical point of view we project the characters of \ the $%
R^{2}$\ even theory which are closed under the full modular group onto the
final theory which is closed only under the $\Gamma _{0}(2)$ subgroup.

The characters of the theory are reduced to $\chi _{(i,s)}^{+}$, with the
result that only these linear combinations can be factorized in terms of the
$c=\frac{3}{2}$ \ and $c=\frac{1}{2}$ theory. This is due to the parity
selection rule.

\subsection{The characters of the untwisted sector}

The characters of the untwisted sector are given in terms of the CFT
components $\tilde{\chi}_{(\left( i,s\right) ,\tilde{f})}$ and $\tilde{\chi}%
_{(s)},$ which are defined below as
\begin{eqnarray}
\tilde{\chi}_{((0,s),0)}(w|\tau ) &=&\bar{\chi}_{0}\chi _{(0,s)}^{MR}(w|\tau
) \\
\tilde{\chi}_{((0,s),1)}(w|\tau ) &=&\bar{\chi}_{\frac{1}{2}}\chi
_{(2,s)}^{MR}(w|\tau ) \\
\tilde{\chi}_{((1,s),0)}(w|\tau ) &=&\bar{\chi}_{0}\chi _{(2,s)}^{MR}(w|\tau
) \\
\tilde{\chi}_{((1,s),1)}(w|\tau ) &=&\bar{\chi}_{\frac{1}{2}}\chi
_{(0,s)}^{MR}(w|\tau ) \\
\ \ \tilde{\chi}_{(s)}(w|\tau ) &=&\bar{\chi}_{\frac{1}{16}}\chi
_{(1,s)}^{MR}(w|\tau )  \nonumber
\end{eqnarray}
As for the twisted sector we can define the symmetrized and antisymmetrized
linear combinations which can be rewritten as
\begin{eqnarray}
\tilde{\chi}_{(0,s)}^{+}(w|\tau ) &=&\bar{\chi}_{0}\chi _{(0,s)}^{MR}(w|\tau
)+\bar{\chi}_{\frac{1}{2}}\chi _{(2,s)}^{MR}(w|\tau ) \\
\tilde{\chi}_{(1,s)}^{+}(w|\tau ) &=&\bar{\chi}_{0}\chi _{(2,s)}^{MR}(w|\tau
)+\bar{\chi}_{\frac{1}{2}}\chi _{(0,s)}^{MR}(w|\tau ) \\
\tilde{\chi}_{(0,s)}^{-}(w|\tau ) &=&\bar{\chi}_{0}\chi _{(0,s)}^{MR}(w|\tau
)-\bar{\chi}_{\frac{1}{2}}\chi _{(2,s)}^{MR}(w|\tau ) \\
\tilde{\chi}_{(1,s)}^{-}(w|\tau ) &=&\bar{\chi}_{0}\chi _{(2,s)}^{MR}(w|\tau
)-\bar{\chi}_{\frac{1}{2}}\chi _{(0,s)}^{MR}(w|\tau )
\end{eqnarray}
Now we can compare the characters of the TM with the Halperin ones $\chi
_{(f,s)}^{H}(w|\tau )$ and obtain the following identifications
\begin{eqnarray}
\tilde{\chi}_{(0,s)}^{+}(w|\tau ) &=&\chi _{(1,s)}^{H}(w|\tau )  \label{fusi}
\\
\tilde{\chi}_{(1,s)}^{+}(w|\tau ) &=&\chi _{(2,s)}^{H}(w|\tau ) \\
\tilde{\chi}_{(s)}(w|\tau ) &=&\chi _{(3,s)}^{H}(w|\tau )=\chi
_{(4,s)}^{H}(w|\tau )  \label{charhalp}
\end{eqnarray}
We can see that the characters of the ($p+2$,$p+2$,$p$) Halperin model are
linear combinations of the TM characters in the untwisted sector. But we
should notice a very peculiar phenomenon embodied in eq.(\ref{charhalp})
where one character of the TM is identified with $2$ characters of the H
model. Because of that the degeneracy of the ground state on the torus for
this subset of characters is reduced from $4(p+1)$ to $3(p+1)$. That may be
an indication of the transition from the Abelian statistics of the H model
to the non-Abelian one of the TM. Naturally such an interesting issue needs
to be further studied to be completely clarified.

The fields content and the fusion rules for the conformal family can be
deduced from the properties of the characters under modular transformations
(see App. B). Using the results given in App. B we verify that the
characters are closed under $SL(2,Z)$ for $p$ even or under $\Gamma _{0}(2)$
for $p$ odd. It is important to notice that there is a further symmetry,
generated by the magnetic translations \cite{cgm5}\cite{Traslmagn}. The two
generators of this group are the shifts $w\rightarrow w+1$ which implies
that the electrons have an integer charge and $w\rightarrow w+\tau $ which
is a self-consistency condition for the completeness of the excitations
under the change of the electric potential and corresponds to a transport of
a magnetic flux unit along a cycle of the torus. It is easy to see that the
characters of the TM are a complete basis for this group. \ \

Their explicit derivation and the complete set of fusion rules together with
the issue of non-Abelian statistics will be discussed in details elsewhere
\cite{workinprogress}.

\section{The partition function}

In this section we derive the partition function for the TM and confront it
with the well known function for the Halperin and Moore-Read models.

Starting \ from the diagonal partition function of the mother theory and
using the modular properties of the characters of the twisted and untwisted
sectors given in App. B, we obtain the following partition function:
\begin{equation}
Z(\tau )=\sum_{s=0}^{p}\left( \sum_{k}{\LARGE \ }\left| \tilde{\chi}%
_{k}(0|\tau )\right| ^{2}+\sum_{t}\left| \chi _{t}(0|\tau )\right|
^{2}\right)
\end{equation}
where the index $k$ ($t$) runs on the untwisted (twisted) characters.

This function can also be written in a diagonal form in terms of the
symmetrized and antisymmetrized characters defined in secs.(4.1), (4.2) as:
\begin{equation}
Z(\tau )=\frac{1}{2}\left( Z^{H}(\tau )+Z_{untwist}^{-}(\tau
)+Z_{twist}^{+}(\tau )+Z_{twist}^{-}(\tau )\right)  \label{Z}
\end{equation}
Above $Z^{H}$ is the partition function of the Halperin model\footnote{%
notice that the reduction of the number of characters of the H model
entering the TM (see eq.(\ref{charhalp}) and comments afterwards) does not
affect $Z^{H}$.}
\begin{equation}
Z^{H}(\tau )=\sum_{s=0}^{p}\sum_{i=0}^{4}|\chi _{(i,s)}^{H}(0|\tau )|^{2}
\end{equation}
which can be expressed in terms of our characters as:
\begin{equation}
Z^{H}(\tau )=\sum_{s=0}^{p}2\left| \tilde{\chi}_{(s)}(0|\tau )\right|
^{2}+Z_{untwist}^{+}(0|\tau )
\end{equation}
where
\begin{equation}
Z_{untwist}^{+}(\tau )=\sum_{s=0}^{p}\left( \left| \tilde{\chi}%
_{(0,s)}^{+}(0|\tau )\right| ^{2}+\left| \tilde{\chi}_{(1,s)}^{+}(0|\tau
)\right| ^{2}\right)
\end{equation}
The other terms in eq.(\ref{Z}) are explicitly given by:
\begin{eqnarray}
Z_{untwist}^{-}(\tau ) &=&\sum_{s=0}^{p}\left( \left| \tilde{\chi}%
_{(0,s)}^{-}(0|\tau )\right| ^{2}+\left| \tilde{\chi}_{(1,s)}^{-}(0|\tau
)\right| ^{2}\right) \\
Z_{twist}^{+}(\tau ) &=&\sum_{s=0}^{p}\left( \left| \chi _{(0,s)}^{+}(0|\tau
)\right| ^{2}+\left| \chi _{(1,s)}^{+}(0|\tau )\right| ^{2}\right) \\
Z_{twist}^{-}(\tau ) &=&\sum_{s=0}^{p}\left( \left| \chi _{(0,s)}^{-}(0|\tau
)\right| ^{2}+\left| \chi _{(1,s)}^{-}(0|\tau )\right| ^{2}\right)
\end{eqnarray}

Using the modular transformations properties of the characters given in App.
B it is easy to verify that the modular properties of $Z_{untwist}^{-}$, $%
Z_{twist}^{+}$ and\ $Z_{twist}^{-}$ do not depend on the parity of the flux $%
p$ while for $Z^{H}$ this is not the case.

Under $T$ they transform as:

\begin{center}
\begin{equation}
\left.
\begin{array}{c}
\left\{
\begin{array}{c}
even~p:~~Z^{H}(\tau +1)=Z^{H}(\tau ) \\
odd~p:~~Z^{H}(\tau +2)=Z^{H}(\tau )
\end{array}
\right. ;~~\ \ \ \ \ \ \ ~Z_{untwist}^{-}(\tau +1)=Z_{untwist}^{-}(\tau ) \\
Z_{twist}^{+}(\tau +1)=Z_{twist}^{-}(\tau );~\ \ \ \ \ \ \ \ \ \ \ \ \ \ \
\ \ \ \ \ \ \ \ \ \ ~Z_{twist}^{-}(\tau +1)=Z_{twist}^{+}(\tau )
\end{array}
\right.   \label{tras.modul.T}
\end{equation}
\end{center}

Under $S$ transformation we have:
\begin{eqnarray}
Z^{H}(-\frac{1}{\tau })&=&Z^{H}(\tau );~~~~~~~~
Z_{untwist}^{-}(-\frac{1}{\tau })=Z_{twist}^{+}(\tau ) \\
Z_{twist}^{+}(-\frac{1}{\tau })&=&Z_{untwist}^{-}(\tau
);~~~~~Z_{twist}^{-}(-\frac{1}{\tau })=Z_{twist}^{-}(\tau )
\label{tras.modul.S}
\end{eqnarray}
From the transformations given above it is clear that $Z^{H}$ is modular
invariant for $p$\ even while it is only weak modular invariant, i.e. $%
\Gamma _{0}(2)$ invariant, for $p$ odd.

\subsection{Z$_{2}$-orbifold construction of the Halperin model}

For a CFT on the torus with an internal $Z_{2}$ symmetry it is possible to
construct a modular invariant partition function which takes into account
different boundary conditions along the space and the time coordinates. The
orbifold construction gives rise to a partition function which has the
following structure:
\[
Z(\tau )\equiv \frac{1}{2}\left\{ Z_{\left( P,P\right) }(\tau )+Z_{\left(
P,A\right) }(\tau )+Z_{\left( A,P\right) }(\tau )+Z_{\left( A,A\right)
}(\tau )\right\}
\]
where $P$ ($A$) stays for periodic (antiperiodic) boundary conditions.

The terms appearing above have the following modular transformations
property:
\begin{eqnarray}
T\left( Z_{\left( P,P\right) }\left( \tau \right) \right)  &\propto
&Z_{\left( P,P\right) }\left( \tau \right) ;~~~~~~~~T(Z_{\left( P,A\right)
}\left( \tau \right) )\propto Z_{\left( P,A\right) }\left( \tau
\right)  \\
T(Z_{\left( A,P\right) }\left( \tau \right) ) &\propto &Z_{\left(
A,A\right) }\left( \tau \right);~~~~~~~~T(Z_{\left( A,A\right) }\left(
\tau \right) )\propto Z_{\left( A,P\right) }\left( \tau \right)
\end{eqnarray}
and
\begin{eqnarray}
S\left( Z_{\left( P,P\right) }\left( \tau \right) \right)  &\propto
&Z_{\left( P,P\right) }\left( \tau \right);~~~~~~~~~~~S(Z_{\left(
P,A\right) }\left(
\tau \right) )\propto Z_{\left( A,P\right) }\left( \tau
\right)  \\
S(Z_{\left( A,P\right)}\left( \tau \right) ) &\propto &Z_{\left( P,A\right)
}\left( \tau \right);~~~~~~~~~~~S(Z_{\left( A,A\right) }\left( \tau
\right) )\propto Z_{\left( A,A\right) }\left( \tau \right)
\end{eqnarray}

If we compare the modular transformations $T$ and $S$ of $Z^{H}${\bf , }$%
Z_{untwist}^{-}$, $Z_{twist}^{+}$, $Z_{twist}^{-}$, indicated in eqs.($\ref
{tras.modul.T}$-$\ref{tras.modul.S}$), with the above ones, we can make for $%
p$ even the following identification: {\bf \ }
\begin{eqnarray}
Z^{H}(\tau ) &\equiv &{\bf \ }Z_{\left( P,P\right) }(\tau ) \\
{\bf \ }Z_{untwist}^{-}(\tau ){\bf \ } &\equiv &{\bf \ }Z_{\left( P,A\right)
}(\tau ){\bf .} \\
Z_{twist}^{+}(\tau ) &\equiv &{\bf \ }Z_{\left( A,P\right) }(\tau ){\bf .} \\
{\bf \ }Z_{twist}^{-}(\tau ) &\equiv &{\bf \ }Z_{\left( A,A\right) }(\tau )
\end{eqnarray}

Notice that the two Majorana fermions, introduced in eqs:(\ref{even}-\ref
{odd}) are not completely equivalent because the monodromy conditions of the
wave function select out the twist invariant $\psi $ fermion as the one
associated to the charged sector. That can be also seen in the factorization
of the partition function given in eq.(\ref{Z}) as:
\begin{equation}
Z(\tau )=Z^{MR}(\tau )Z_{\overline{I}}(\tau )
\end{equation}
where $Z^{MR}$ is the modular invariant partition function of the Moore-Read
model, which has central charge $c=3/2$ and $Z_{\overline{I}}$ is the
partition function of the Ising model, which has $c=1/2$. They have the
following expression:
\begin{equation}
Z_{MR}(\tau )=\sum_{s=0}^{p}\left( \left| \chi _{(0,s)}^{MR}(0|\tau )\right|
^{2}+\left| \chi _{(1,s)}^{MR}(0|\tau )\right| ^{2}+\left| \chi
_{(2,s)}^{MR}(0|\tau )\right| ^{2}\right)
\end{equation}
and
\begin{equation}
Z_{\overline{I}}(\tau )=\left| \bar{\chi}_{0}(\tau )\right| ^{2}+\left| \bar{%
\chi}_{\frac{1}{2}}(\tau )\right| ^{2}+\left| \bar{\chi}_{\frac{1}{16}}(\tau
)\right| ^{2}
\end{equation}

For $p$ odd it can be easily seen that $Z$ can be written as:
\begin{equation}
Z(\tau )=\frac{1}{2}\left( Z^{H}(\tau )+Z_{untwist}^{-}(\tau
)+Z_{twist}^{+}(\tau )\right)
\end{equation}
which is the only expression consistent with the parity rule; furthermore
the above partition function is invariant only under $\Gamma _{0}(2)$. For
such a case it is natural to expect a connection between the parity rule and
the modular properties of the characters.

In \cite{INO} it was shown that the H and the HR partition function are
related by a half-flux transport along a cycle of the torus. In a
forthcoming paper \cite{workinprogress} we extend this argument by using the
symmetry generated by magnetic translations generalized to the case of
transport of the flux from a layer to the other.

\section{Conclusions}

In this paper, by using the $m$-reduction procedure, we construct the
characters of the $c=2$ twisted CFT, and studied their modular properties.
The new aspect in our construction is the presence, induced in the daughter
CFT, of discrete symmetries which couple to each other due essentially to
the compositeness of the vertex operators describing the primary fields of
the theory. From a physical point of view a first $Z_{2}$ can be seen as due
to the presence of a current between the layers which flows at a fixed but
arbitrary point of the edge. This current can be due to the presence of a
contact point \cite{Bimonte}. The presence of an interlayer current also for
very weak tunneling has been recently observed in the ($1,1,1$) state\cite
{Experiment}. Then it is natural to interpret the twisted boundary
conditions for the scalar fields as the theoretical description in our
context of such a current. The TM takes into account such an effect for weak
tunneling; on the other hand when there is a uniform strong tunneling the
model should flow to the MR model as discussed before.

There is also a second $Z_{2}$ symmetry which acts on the vertex operators
for the neutral modes as it can be seen in eqs.(19,20). Then the possibility
to select a symmetric or an antisymmetric vertex is related to the
phenomenon of reduction of the number of degrees of freedom. As an
interesting result there is an enhancement of the symmetry, i.e. the $N=2$
superconformal symmetry (for $p=1$ or equivalently $\nu =1/2$) present in
the $c=3/2$ CFT, which has been evidenced in \cite{cgm3}.

Furthermore the uniform strong tunneling phenomenon in bilayer systems has
been also studied with different field theory techniques as the abelian
bosonization ( \cite{priadko},\cite{Cabra}). Even though in the first work
the R.G.-flow is not present, their results are in agreement with the
analysis given in the previous sections.

It is now the correct time to ask ourselves why we can construct two
different CFTs (with $c=2$) which seem to be relevant for the bilayer system
before the uniform tunneling takes place. We are referring to the original H
model and to the TM presented in this paper (secs. 4 and 5).

The two models differ for the boundary conditions imposed on the $m=2$
scalar fields, of which one ($m-1$) do satisfy twisted boundary conditions
(see eq.(11)) for the TM case. Then it is natural to view that as a
description of a particle interaction in the presence of localized
impurities. Further such an interaction could account for a current flow
between the layers also in the absence of uniform tunneling. All that is in
agreement with recent experimental observations \cite{Experiment}; other
theoretical analysis in such a context have been already presented in \cite
{goldstone}. \ Then a coherent superposition of interlayer interactions
could drive the system to a more symmetric phase in which the two layers are
indistinguishable, i.e. the MR state. That is consistent with the analysis
given in \cite{cgm3} for the brane physics where the tunneling phenomenon,
interpreted as a tachyon condensation process, allows for bound states of
two D2 branes system. Naturally such an interpretation needs to be verified
by further analysis.

We should point out that while in the H model the statistics is certainly
Abelian, in the TM the statistics appears to be non-Abelian, as for the MR
model (see eq.(\ref{charhalp}) and comments afterwards). Such a transition
from an Abelian to a non-Abelian statistics seems to be due to the
decoupling of the spin structures, i.e. the presence of two independent
inequivalent Majorana fermions and the breaking of the symmetry which
exchanges them.

The observed reduction of the independent characters in the untwisted sector
of the TM, constructed in sec. 5, requires an accurate analysis of the zero
modes of the fermionic fields. We just observe that by introducing the well
known operator $(-1)^{F}$, which makes possible the projection on the even
or odd fermionic number, one would be able to distinguish between the order
operator $\sigma $ from the disorder operator $\mu $ in the Ising model. For
the H model the states of both parity are on the same footing, then the two
(independent) orthogonal characters $\chi _{\frac{1}{16}}$ are available.
That justifies the presence of the two characters $\chi _{(3,s)}^{H}$ and $%
\chi _{(4,s)}^{H}$ as independent ones. Instead for the orbifold model the
modular invariance requires a GSO-like projection for both the Ising models
\cite{Ginsparg}, reducing then the characters entering the $Z_{\left(
P,P\right) }$ partition function to $3(p+1)$ (see eq.(\ref{untwisted})). An
interesting investigation on related issues has been recently presented in
\cite{Georgiev}.

Then the difference between the Halperin model and the TM is that while in
the first one the fundamental particles are Dirac fermions with a well
defined layer index in the second one they are expressed in terms of
symmetric $\psi $ and antisymmetric $\bar{\psi}$ fields, which are a
superposition of states belonging to different layers. This basis is
relevant for the TM because the twisted boundary conditions of the boson
distinguish the two types of fermions also when the tunneling is absent.

{\bf Acknowledgments} - We thank M. Huerta for useful comments and
discussions. We also thank A. Bergere and V. Pasquier for their interest to
this work and for the hospitality extended to one of us [GM] at the Service
de Physique Theorique, Saclay-F.

{\large {\bf Appendix A}}

{\large {\bf The twisted sector of the c=m TM }}

The twisted sector of the TM is obtained by $m$-reduction from the $U(1)_{a}$
theory, with central charge $c=1$, and radius of compactification for the
free boson $R^{2}=pm+2\equiv 2a$ with $pm$ even and $m$ prime. The theory so
obtained has central charge $c=m$.

Furthermore the $m$-reduction procedure allows us to construct the conformal
blocks of the TM and to recognize the conformal dimension of the primary
fields they represent. It is interesting to notice that their transformation
properties under the modular group not only give us information about the
irreducibility of the representations but at the same time gives us the
fusion rules. In analogy with the construction made on the plane, we can
express the characters of the daughter theory in terms of the characters of
the mother one.

In the daughter theory the conformal dimensions of the primary fields in the
twisted sector and the corresponding characters are respectively given by:
\begin{equation}
h_{\tilde{f}}^{q}=\frac{q^{2}}{2(pm+2)m}+\frac{m^{2}-1}{24m}+\frac{\tilde{f}%
}{m}  \label{dim}
\end{equation}
\begin{equation}
\chi _{(q,\tilde{f})}(w|\tau )=\frac{1}{m}\sum_{j=0}^{m-1}e^{-\frac{2\pi i}{m%
}j(\tilde{f}+\frac{q^{2}}{2(pm+2)}-\frac{1}{24})}K_{q}^{\left( a\right) }(w|%
\frac{\tau +j}{m})  \label{carattere}
\end{equation}
where $q=ms+2b(1-\delta _{s,p})+i$ with \ $\left[ s,2b,i\right] =\left[
(0,.,p),(0,.,m-1),(0,1)\right] $ ; $\tilde{f}=0,...,m-1.$ Above the
characters of the $U(1)_{a}$ mother theory are explicitly given by:
\begin{equation}
K_{q}(w|\tau )=\frac{1}{\eta (\tau )}\Theta \left[
\begin{array}{c}
\frac{q}{(pm+2)} \\
0
\end{array}
\right] \left( aw|2a\tau \right)   \label{Chid}
\end{equation}
corresponding to the primary fields with conformal dimensions $h_{q}=\frac{%
q^{2}}{2(pm+2)}.$ In the daughter theory the charged sector is described by
a $U(1)_{ma}$ with $c=1$ and radius of compactification of the free boson
given by $R_{X}^{2}=\left( pm+2\right) /m.$ Its conformal blocks can be
expressed as
\begin{equation}
K_{2al+q}(w|\tau )=\frac{1}{\eta (\tau )}\Theta \left[
\begin{array}{c}
\frac{2al+q}{m(pm+2)} \\
0
\end{array}
\right] \left( amw|2am\tau \right)
\end{equation}
where $l=0,...,m-1$.

Also the neutral sector is described by a CFT\ with central charge $c=m-1$.
Of course the two sectors are not independent as it clearly appears from the
following decomposition of the characters of the TM:
\begin{equation}
\left. \chi _{(q,\tilde{f})}(w|\tau )=\sum_{l=0}^{m-1}N_{{\large (}l,(q,%
\tilde{f}){\large )}}^{(p)}(\tau )K_{2al+q}(w|\tau )\right.   \label{Chrel}
\end{equation}
with the neutral characters $N_{(l,(q,\tilde{f}))}^{(p)}$ given by:
\begin{equation}
N_{{\large (}l,(q,\tilde{f}){\large )}}^{(p)}(\tau )=\frac{1}{m}%
\sum_{j=0}^{m-1}e^{-\frac{2\pi i}{m}j(\tilde{f}+\frac{q^{2}}{4a}-\frac{1}{24}%
-\frac{\left( 2al+q\right) ^{2}}{4a})}\frac{\eta (\tau )}{\eta (\frac{\tau +j%
}{m})}
\end{equation}
The above decomposition in eq.(\ref{Chrel}) even do lengthy can be easily
obtained by using in eq.(\ref{carattere}) the following identity, relating $%
\Theta $ with different arguments:
\begin{equation}
\Theta \left[
\begin{array}{c}
\frac{q}{2a} \\
0
\end{array}
\right] \left( aw|2a\frac{\tau +j}{m}\right) =\sum_{l=0}^{m-1}e^{2\pi ij%
\frac{\left( 2al+q\right) ^{2}}{4ma}}\Theta \left[
\begin{array}{c}
\frac{2al+q}{2ma} \\
0
\end{array}
\right] \left( maw|2ma\tau \right)
\end{equation}
where $q=0,...,mp+1$.

We must observe that the neutral part does not depend on $p$, \ that is on
the flux attached to the charged component, the only dependence on $p$ being
the way the characters of the neutral part couple to the charged ones in the
TM. Such an independence is an immediate consequence of the following
relation between neutral characters:
\begin{equation}
\left. N_{{\large (}l,(q,\tilde{f}){\large )}}^{\left( p\right) }(\tau )=N_{%
{\large (}l^{\prime },(q^{\prime },\ \tilde{f}^{\prime }{\large ))}}^{\left(
0\right) }(\tau )\right.   \label{neutro}
\end{equation}
where $l^{\prime }=l+b(1-\delta _{s,p})$, $\ q^{\prime }=i$, $\ \tilde{f}%
^{\prime }=\tilde{f}-\frac{^{pml^{2}}}{2}-b(1-\delta _{s,p})(b+i)$, $\ \
l=0,...,m-1$ and $q=ms+2b(1-\delta _{s,p})+i$.

Furthermore by using eq.$\left( \ref{neutro}\right) $ we get for the
characters of the TM the following expression:
\begin{equation}
\chi _{(q,\tilde{f})}(w|\tau )=\sum_{l=0}^{m-1}N_{{\large (l^{\prime
},(q^{\prime },\tilde{f}^{\prime }))}}(\tau )K_{2al+q}(w|\tau )
\label{F(m,p)1}
\end{equation}
where we have defined $N_{(l^{\prime },(q^{\prime },\ \tilde{f}^{\prime
}))}=N_{(l^{\prime },(q^{\prime },\ \tilde{f}^{\prime }))}^{\left( 0\right)
},$ showing the way the neutral and charged components couple together. From
the above equation it appears that such a \ coupling depends on the parity
of $p$.

{\large {\bf Decomposition of the neutral component N}}

In section 3 for the theory on the plane we noticed that the neutral
components $\psi _{1}(z)$ appearing in eq.(\ref{par.1}) satisfy the
parafermionic algebra for $SU(2)$ at level $m$ \cite{cgm2}.

Of course the neutral degrees of freedom content of the TM corresponding to
a central charge $c=m-1$, is richer than that of the parafermionic theory
which has $c=2\left( m-1\right) /\left( m+2\right) $. In fact the neutral
contribution to the conformal dimensions in eq.(\ref{dim}) splits into two
terms:
\begin{eqnarray}
h_{\lambda ,l}^{i} &=&\frac{\lambda (m-\lambda )}{2(m+2)}+\frac{(\frac{%
\lambda }{2}-l-\frac{i}{2})(l+\frac{i}{2}+\frac{\lambda }{2})}{m} \\
\bar{h}_{\lambda ,\tilde{f}}^{i} &=&\frac{m^{2}-1}{24m}-\frac{\lambda
(m-\lambda )}{2(m+2)}+\frac{(\frac{i}{2}-\frac{\lambda }{2})(\frac{i}{2}+%
\frac{\lambda }{2})}{m}+\frac{\tilde{f}}{m}
\end{eqnarray}
where $h_{\lambda ,l}^{i}$ is the dimension of the primary fields in the
parafermionic theory while $\bar{h}_{\lambda ,\tilde{f}}^{i}$ is the
corresponding one for the coset CFT which has central charge $\bar{c}%
=m\left( m-1\right) /\left( m+2\right) $. On the torus that shows up
immediately in the following decomposition of the characters of the neutral
part $N_{(l,(i,\tilde{f}))}$ in terms of those of the parafermionic theory $%
C_{(\lambda ,q)}^{\left( m\right) }$:
\begin{equation}
N_{{\large (}l,(i,\tilde{f}){\large )}}(\tau )=\sum_{i+\lambda =0~{\em mod}%
2~,\lambda =0}^{m}{\large C}_{(\lambda ,2l+i)}^{\left( m\right) }(\tau )b_{%
{\large (}\lambda ,(i,\tilde{f}){\large )}}^{\left( m\right) }(\tau )
\label{eq.11}
\end{equation}
with $q=0,...,2m-1$ and $\lambda =0,...,m$.

The characters $C_{(\lambda ,q)}^{\left( m\right) }$ are written in terms of
the string functions $c_{(\lambda ,q)}$ as
\begin{equation}
C_{(\lambda ,q)}^{\left( m\right) }(\tau )=\eta (\tau )c_{(\lambda ,q)}(\tau
)
\end{equation}

and $b_{(\lambda ,(i,\tilde{f}))}^{\left( m\right) }$ are the branching
functions of the neutral theory coset.

For the special $m=2$ case the neutral sector has central charge $c=1$ and
the parafermionic theory coincides with the theory of a free Majorana
fermion with central charge $c=1/2$. Its characters are explicitly given by:
\begin{eqnarray}
\chi _{0}(\tau ) &=&\frac{1}{2}\left( \sqrt{\frac{\Theta _{3}\left( 0|\tau
\right) }{\eta (\tau )}}+\sqrt{\frac{\Theta _{4}\left( 0|\tau \right) }{\eta
(\tau )}}\right)  \\
\chi _{\frac{1}{2}}(\tau ) &=&\frac{1}{2}\left( \sqrt{\frac{\Theta
_{3}\left( 0|\tau \right) }{\eta (\tau )}}-\sqrt{\frac{\Theta _{4}\left(
0|\tau \right) }{\eta (\tau )}}\right)  \\
\chi _{\frac{1}{16}}(\tau ) &=&\sqrt{\frac{\Theta _{2}\left( 0|\tau \right)
}{2\eta (\tau )}}
\end{eqnarray}
For this special case the coset theory is still a free fermion theory with
central charge $c=1/2$ and the characters are explicitly given by:
\begin{eqnarray}
b_{{\large (}0,(0,0){\large )}}^{\left( 2\right) }(\tau ) &=&\bar{\chi}_{%
\frac{1}{16}}(\tau );~~b_{{\large (}2,(0,0){\large )}}^{\left( 2\right)
}(\tau )=0;~~b_{{\large (}0,(0,1){\large )}}^{\left( 2\right) }(\tau )=0
\nonumber \\
b_{{\large (}2,(0,1){\large )}}^{\left( 2\right) }(\tau ) &=&\bar{\chi}_{%
\frac{1}{16}}(\tau );~~b_{{\large (}1,(1,0){\large )}}^{\left( 2\right)
}(\tau )=\bar{\chi}_{1}(\tau );~~~b_{{\large (}1,(1,1){\large )}}^{\left(
2\right) }(\tau )=\bar{\chi}_{\frac{1}{2}}(\tau )  \label{le b}
\end{eqnarray}

{\large {\bf Appendix B}}

{\large {\bf Modular transformations for the TM with $c=2$}}

The modular transformations of the characters of the $c=2$ TM are given
below.

Under $T$ ($\tau \rightarrow \tau +1$) the characters of the twisted sector
defined in sec. 4.1 transform for any $p$ as
\begin{eqnarray}
\chi _{(0,s)}^{+}(w|\tau +1)&=&e^{\frac{\pi is^{2}}{p+1}-\frac{\pi i}{24}%
}\chi _{(0,s)}^{-}(w|\tau )~~ \\
\chi _{(1,s)}^{+}(w|\tau +1)&=&e^{\frac{\pi i\left( s+\frac{1}{2}\right) ^{2}%
}{p+1}-\frac{\pi i}{24}}\chi _{(1,s)}^{-}(w|\tau ) \\
\chi _{(0,s)}^{-}(w|\tau +1)&=&e^{\frac{\pi is^{2}}{p+1}-\frac{\pi i}{24}%
}\chi _{(0,s)}^{+}(w|\tau )~~ \\
\chi _{(1,s)}^{-}(w|\tau +1)&=&e^{\frac{\pi i\left( s+\frac{1}{2}\right) ^{2}%
}{p+1}-\frac{\pi i}{24}}\chi _{(1,s)}^{+}(w|\tau ) \\
\tilde{\chi}_{(i,s)}^{-}(w|\tau +1)&=&\left( -1\right) ^{\left( p+1\right)
i}e^{\frac{\pi is^{2}}{p+1}-\frac{\pi i}{6}}\tilde{\chi}_{(i,s)}^{-}(w|\tau )
\end{eqnarray}
Instead the characters of the untwisted sector $\tilde{\chi}_{(i,s)}^{+}$
and $\tilde{\chi}_{(s)},$ defined in sec. 4.2, are closed under $T$ only if $%
p$ is even:
\begin{eqnarray}
~\tilde{\chi}_{(i,s)}^{+}(w|\tau +1) &=&\left( -1\right) ^{i}e^{\frac{\pi
is^{2}}{p+1}-\frac{\pi i}{6}}\tilde{\chi}_{(i,s)}^{+}(w|\tau ) \\
\tilde{\chi}_{(s)}(w|\tau +1) &=&e^{\frac{\pi i\left( s+\frac{1}{2}\right)
^{2}}{p+1}+\frac{\pi i}{12}}\tilde{\chi}_{(s)}(w|\tau )
\end{eqnarray}

while they are closed under $T^{2}$ for $p$ odd:
\begin{eqnarray}
\tilde{\chi}_{(i,s)}^{+}(w|\tau +2) &=&e^{\frac{2\pi is^{2}}{p+1}-\frac{\pi i%
}{3}}\tilde{\chi}_{(i,s)}^{+}(w|\tau ) \\
\tilde{\chi}_{(s)}(w|\tau +2) &=&e^{\frac{2\pi i\left( s+\frac{1}{2}\right)
^{2}}{p+1}+\frac{\pi i}{6}}\tilde{\chi}_{(s)}(w|\tau )
\end{eqnarray}
Under $S$ ($\tau \rightarrow -1/\tau $) we have the following modular
properties for any $p$:
\begin{eqnarray}
\tilde{\chi}_{(i,s)}^{+}(\frac{w}{\tau }|-\frac{1}{\tau })&=&\sum_{i^{\prime
}=0}^{1}\sum_{s^{\prime }=0}^{p}\frac{e^{\frac{2\pi iss^{\prime }}{p+1}}}{2%
\sqrt{(p+1)}}\tilde{\chi}_{(i^{\prime },s^{\prime })}^{+}(w|\tau
)+\sum_{s^{\prime }=0}^{p}\frac{\left( -1\right) ^{i}e^{\frac{2\pi is\left(
s^{\prime }+\frac{1}{2}\right) }{p+1}}}{\sqrt{(p+1)}}\tilde{\chi}%
_{(s^{\prime })}(w|\tau ) \\
\tilde{\chi}_{(s)}(\frac{w}{\tau }|-\frac{1}{\tau })&=&\sum_{i^{\prime
}=0}^{1}\sum_{s^{\prime }=0}^{p}\frac{\left( -1\right) ^{i^{\prime }}e^{%
\frac{2\pi is^{\prime }\left( s+\frac{1}{2}\right) }{p+1}}}{2\sqrt{(p+1)}}%
\tilde{\chi}_{(i^{\prime },s^{\prime })}^{+}(w|\tau ) \\
\tilde{\chi}_{(i,s)}^{-}(\frac{w}{\tau }|-\frac{1}{\tau })&=&\sum_{i^{\prime
}=0}^{1}\sum_{s^{\prime }=0}^{p}\frac{\left( -1\right) ^{i^{\prime }i}e^{%
\frac{2\pi iss^{\prime }}{p+1}}}{\sqrt{2(p+1)}}\chi _{(i^{\prime },s^{\prime
})}^{+}(w|\tau ) \\
\chi _{(i,s)}^{+}(\frac{w}{\tau }|-\frac{1}{\tau })&=&\sum_{i^{^{\prime
}}=0}^{1}\sum_{s^{\prime }=0}^{p}\frac{\left( -1\right) ^{i^{\prime }i}e^{%
\frac{2\pi iss^{\prime }}{p+1}}}{\sqrt{2(p+1)}}\tilde{\chi}_{(i^{\prime
},s^{\prime })}^{-}(w|\tau )
\end{eqnarray}
On the other hand the modular transformations for the characters $\chi
_{(i,s)}^{-}$ depend on the parity of $p$:

for $p$ even
\begin{equation}
\chi _{(i,s)}^{-}(\frac{w}{\tau }|-\frac{1}{\tau })=\sum_{i^{\prime
}=0}^{1}\sum_{s^{\prime }=0}^{p}\frac{\left( 1-\delta _{i,i^{\prime
}}\right) e^{\frac{2\pi i\left( s+\frac{i}{2}\right) \left( s^{\prime }+%
\frac{i^{\prime }}{2}\right) }{p+1}}}{\sqrt{\left( p+1\right) }}\chi
_{(i^{\prime },s^{\prime })}^{-}(w|\tau )
\end{equation}

for $p$ odd
\begin{equation}
\chi _{(i,s)}^{-}(\frac{w}{\tau }|-\frac{1}{\tau })=\sum_{i^{\prime
}=0}^{1}\sum_{s^{\prime }=0}^{p}\frac{\delta _{i,i^{\prime }}e^{\frac{2\pi
i\left( s+\frac{i}{2}\right) \left( s^{\prime }+\frac{i^{\prime }}{2}\right)
}{p+1}}}{\sqrt{\left( p+1\right) }}\chi _{(i^{\prime },s^{\prime
})}^{-}(w|\tau )
\end{equation}

\end{document}